\documentclass[12pt]{article}
\usepackage{graphicx}
\usepackage{url}
\usepackage{cite}
\usepackage[margin=1.25 in]{geometry}
\usepackage[colorlinks = true, linkcolor = blue, urlcolor = blue,
      citecolor = blue, anchorcolor = blue]{hyperref}

\newcommand\pubnumber{} 
\newcommand\pubdate{November 2022}

\def\SLAC{SLAC,
    Stanford University, Menlo Park, California 94025 USA}
\def\doeack{\footnote{Work supported by the US Department of Energy,
                     contract DE--AC02--76SF00515.}}

\def\Title#1{\begin{center} {\Large #1 } \end{center}}
\def\Author#1{\begin{center}{ \sc #1} \end{center}}

\newcommand\pubblock{\rightline{\begin{tabular}{l} \pubnumber\\
         \pubdate \end{tabular}}}
\newenvironment{Abstract}{\begin{quotation} \begin{center}
                       ABSTRACT
     \end{center}\bigskip  }{\end{quotation}}

\def\beq{\begin{equation}}
\def\eeq#1{\label{#1}\end{equation}}
\def\eeqn{\end{equation}}

\newenvironment{Eqnarray}%
   {\arraycolsep 0.14em\begin{eqnarray}}{\end{eqnarray}}
\def\beqa{\begin{Eqnarray}}
\def\eeqa#1{\label{#1}\end{Eqnarray}}
\def\eeqan{\end{Eqnarray}}

\let\bar=\overbar

\def\lsim{\mathrel{\raise.3ex\hbox{$<$\kern-.75em\lower1ex\hbox{$\sim$}}}}
\def\gsim{\mathrel{\raise.3ex\hbox{$>$\kern-.75em\lower1ex\hbox{$\sim$}}}}

\def\del{\partial}
\def\Dslash{\not{\hbox{\kern-4pt $D$}}}
\def\dslash{\not{\hbox{\kern-2pt $\del$}}}

\def\Dlr{\mathrel{\raise1.5ex\hbox{$\leftrightarrow$\kern-1em\lower1.5ex\hbox{$D$}}}}

\def\ee{e^+e^-}

\def\msb{{\bar{\scriptsize M \kern -1pt S}}}

\def\drb{{\bar{\scriptsize D \kern -1pt R}}}

\makeatletter
\def\section{\@startsection{section}{0}{\z@}{5.5ex plus .5ex minus
 1.5ex}{2.3ex plus .2ex}{\large\bf}}
\def\subsection{\@startsection{subsection}{1}{\z@}{3.5ex plus .5ex minus
 1.5ex}{1.3ex plus .2ex}{\normalsize\bf}}
\def\subsubsection{\@startsection{subsubsection}{2}{\z@}{-3.5ex plus
-1ex minus  -.2ex}{2.3ex plus .2ex}{\normalsize\sl}}

\renewcommand{\@makecaption}[2]{%
   \vskip 10pt
   \setbox\@tempboxa\hbox{\small #1: #2}
   \ifdim \wd\@tempboxa >\hsize     
       \small #1: #2\par          
     \else                        
       \hbox to\hsize{\hfil\box\@tempboxa\hfil}
   \fi}

\makeatother

\begin{document}
\begin{titlepage}
\pubblock

\vfill
\Title{Elementary Particle Physics Vision for EPP2024}
\vfill
\Author{ Michael E. Peskin\doeack}
 \medskip
\begin{center} 

  \SLAC 
\end{center}
\vfill
\begin{Abstract}
In the fall of 2022, the decadal survey committee on Elementary
Particle Physics  of the US National Academies
requested 2000 word Vision Papers, giving personal
interpretations of the results of the Snowmass 2021 study and the
future of the field.  This is my
contribution, which emphasizes the central role  of the Higgs boson and
its associated mysteries.
\end{Abstract} 

\vfill
\end{titlepage}

\hbox to \hsize{\null}

\newpage

\def\thefootnote{\fnsymbol{footnote}}
\newpage
\setcounter{page}{1}

\setcounter{footnote}{0}

\begin{large}\begin{center}
    {\bf Elementary Particle Physics Vision for EPP2024~\cite{EPP2024}} \end{center}\end{large}

\bigskip

Particle physics has expanded its scope to include a wide range of
topics from very low to very high energies.  Its core questions,
though, remain the same: to identify the fundamental interactions that
govern the universe, and to demonstrate that these are described by
well-motivated equations of motion.  Today, we know that these
equations include the Standard Model, either as a fundamental or an
effective low-energy description.   The Standard Model is tested to a
remarkable level.  If one includes neutrino Yukawa couplings (an
oversight of the founders), there are arguably no deviations seen in
our measurements of the strong, weak, and electromagnetic
interactions.  But conceptually the Standard Model is far from
perfect. It contains a large number of free parameters whose values it
does not---and cannot---explain.  And, it provides a shaky and
unstable foundation for the pursuit of additional laws of
nature required to solve the mysteries it leaves open.

The foundations of the Standard Model divide neatly into two parts,
with very different status.  On one side, there is the principle of
local gauge invariance.   This is a very beautiful concept that
completely determines the structure of the couplings of fermions and
gauge bosons in terms of three fundamental coupling constants,
corresponding to the three Standard Model gauge groups $SU(3)\times
SU(2)\times U(1)$.  The structure is highly constrained, and its
validity has been dramatically confirmed by experiment, from
low-energy tests of QED through the muon and electron $(g-2)$ and the
spectra of hydrogen and positronium~\cite{Kinoshita},
through the direct and highly
redundant measurements of the gauge couplings at the $Z$
resonance~\cite{LEP},  to
the measurements of $W$ and $Z$ boson production and multiple gluon
radiation at the LHC~\cite{LHC}. All of the experimental successes of the
Standard Model are tests of the consequences
of the gauge principle.

On the other side, there is the Higgs field and its associated
couplings.  Here one might hope that there would be a ``Higgs
principle'' that explains the values and patterns of these couplings.
Today, we have no principles at all. The Standard Model Higgs
couplings are unrestricted---and, because they are parameter inputs of
a renormalizable quantum field theory, they are uncalculable.
Apparent predictions, such as  the statement that CP violation is
largest in the 3rd generation, are consequences of the gauge
structure.   The most central question is that of why electroweak
symmetry is spontaneously broken at all.  In the Standard Model, this
is an arbitrary  choice, with a particular parameter ($\mu^2$) being
chosen to be negative
without explanation.

These considerations put a premium on improving our knowledge of the
Higgs field.  At the moment, we know only one Higgs boson.  But models
that give a dynamical explanation for the Higgs  spontaneous symmetry
breaking predict additional particles.  These include additional
scalars---an extended Higgs sector---and other particles whose
interaction with the scalar  field(s) generate the Higgs potential.
These particles have consequences at currently accessible energies, both
in flavor physics and in the couplings of the Higgs
boson.  We must  take the opportunities we have now to
discover concrete deviations from the Standard Model at current
energies.   It will be our task in the future to explore energies
beyond LHC for new fundamental interactions connected to the Higgs
field, and for this we must  now to develop the new accelerator
technologies that will get us there. 

Many other questions are often cited as key mysteries of particle
physics.  But most of these are also, ultimately, questions
about the Higgs boson and Higgs fields more generally. 

The most obvious question about the Standard Model, one which has been
with us since the discovery of the muon, is the origin of the mass
spectrum of quarks and leptons.  We now have direct experimental
evidence from the LHC, at least for $t$, $b$, and $\tau$, that the
masses of Standard Model fermions are mainly due to their couplings to
the 125 GeV Higgs boson~\cite{HiggsatLHC}.  This does not explain the masses of these
particles; it only pushes the mystery back one level. Similarly, weak
interaction mixing and CP violation in heavy quark decays can be
complete accounted for by the more fundamental matrix of Higgs boson
couplings~\cite{BFactories}. There is a long history of attempts to give a dynamical
explanation of this data, beginning with Harold Fritzsch's 1977
paper~\cite{Fritzsch}. However, despite many original ideas, the origin of the fermion
masses and couplings has only gotten murkier as our knowledge of the
masses and mixings has become more precise.   The emphasis in flavor
physics has now shifted to the search for flavor anomalies such as
lepton non-universality.   These are specifically outside the
framework of the Standard Model and so would provide evidence for new
fundamental interactions. The search for these effects has high
importance. On the other hand, such anomalies must be due to new
interactions at the TeV scale that play into the question of the
determination of the Higgs Yukawa couplings. At best, we will only get
some clues from the presence of anomalies.  We will need to go to high
energies with colliders to
learn the underlying story.

The generation of the cosmic baryon-antibaryon asymmetry
requires a new source of CP violation beyond that in the Standard Model. 
Phases (aside from the strongly constrained
$\theta$ parameter of QCD) enter a fundamental Lagrangian only in
fermion-scalar couplings. 
So this question is again a question about the Higgs boson, or about
an extended Higgs sector.  In the leptogenesis model, the  new
Higgs bosons are those that give masses and mixings to the heavy neutrinos of
seesaw models; it will be a long time before we test such models. But
in other models,
the new Higgs interactions could be at a mass scale
accessible to colliders.  These provide
another important goal for colliders at LHC and higher energies. 

Many questions are asked about neutrinos, but these more reflect our
still-imperfect knowledge of neutrinos rather than opportunities to go
beyond the Standard Model.  We do need to clarify the neutrino mass
ordering and demonstrate CP violation in the neutrino mass matrix. To
ask more fundamental questions, we need to know how the neutrino
masses are generated.  By the gauge principle for the electroweak
interactions, neutrino masses must come from the Higgs field.  The
crucial question is whether the neutrino masses are Dirac, with a very
small value of the Yukawa coupling, or Majorana, from a seesaw with a
heavy right-handed neutrino. This question is not addressed by
long-baseline neutrino experiments, but it lends importance to
searches for neutrinoless double beta decay.  Other commonly asked
questions seem to have small importance.  There is no evidence for
low-mass sterile neutrinos, given the uncertainties on cross sections
from low-energy QCD, and the motivation for additional searches is ad hoc.
Other beyond-Standard-Model effects, such as flavor-changing higher
dimension interactions, will remain more strongly constrained from
measurements on charged leptons, which by gauge symmetry must appear
in these operators,  than from neutrino interactions. 

The nature of dark matter is a major question whose solution is
necessarily outside the Standard Model, and which need not be
connected to the Higgs.  It is very important to find direct evidence
for the particle nature of dark matter and, through this discovery, to
gain insight into its nature.  The idea that dark matter is a particle
with mass at the Higgs boson mass scale is now strongly challenged by
experiment.  This has motivated new searches for dark matter particles
in a lower-mass region that is accessible at current accelerators or
even benchtop experiments.  Current discussions, though, leave out the
possibility of dark matter particles in the TeV mass range that have
higher annihilation rates by being more strongly coupled to the Higgs
sector than previously emphasized candidates such as neutrinos or
sleptons.  Such heavier WIMPs---together with axions, which are also
particles of the Higgs
sector---are the best-motivated dark matter candidates today.

Models of flavor, baryogenesis, and neutrino mass all are built on
hypotheses for the Higgs sector.  Because we do not yet have a Higgs
principle, these models cannot be predictive.  These models rely on
specific but arbitrary choices for the unconstrained scalar
couplings. Until we have a more fundamental understanding of the
nature of the Higgs sector, it will be difficult to make real progress
on any of these issues.  To gain that understanding, we need
experiments that directly
address the nature of the Higgs field.

This statement applies also to our ultimate goal, the unification of
all interactions.  Both dark matter and the mysteries of the Higgs
imply that there are further fundamental forces beyond those of the
Standard Model.  Einstein tried to form a unified theory of nature
using only gravity and electromagnetism, and we now know that that
quest was doomed.  We are in a similar situation today.  We know that
there are additional interactions that are needed to make a global
theory of nature.
We need experiments that will give us knowledge of them. 

This experimental program is a major focus of the Energy Frontier
report for Snowmass 2021~\cite{EFreport}, and the conclusions of that report
especially merit attention.
This program must proceed in three phases.

First, we must continue the search at the LHC for new particles
that could give evidence of a Higgs sector broader than that of the
Standard Model.  Though there is no such evidence yet, the
capabilities of the LHC are hardly exhausted.  These capabilities
will be extended 
in the high-luminosity stage of the LHC~\cite{LHCnewparticles}.

Second, we need to search for evidence of a broader Higgs sector from
detailed 
studies of the known Higgs boson.   This is the most obvious
place to learn about the Higgs sector, but current experiments have not yet
reached the required level of precision.  It is well documented that $\ee$ Higgs
factories have the capability of measuring the Higgs boson couplings
at the sub-percent level, and that such measurements constitute broad
searches for new physics that complement the LHC results~\cite{ILCreport}.  We have
the technology today to construct an $\ee$  Higgs factory, one that is
affordable with global collaboration.  If no other region will step
forward, the
US should take the lead.

Third, we need to develop technologies for robust and cost-effective
exploration of physics at energies of 10 TeV and above in
parton-parton collisions.  Today, approaches are proposed with proton,
muon, electron, and photon colliders.
We need to bring at least a subset of these to maturity.

The vision for the future of particle physics must acknowledge the
central role of the Higgs field.  The Higgs field is a crucial part of
the Standard Model.  It is our ignorance about this field that keeps
us from solving the remaining mysteries that the Standard Model cannot
address. To make progress, we must remedy this.  We need to make clear
(with apologies to Red Sanders and Vince Lombardi): {\bf ``Higgs isn't
everything; it's the only thing.''}  A vision for particle physics that
is not built on this idea cannot address the most profound questions for our
field or realize its greatest
opportunities.

\end{document}